\documentclass[12pt]{article}
\usepackage[utf8]{inputenc}
\usepackage{amsfonts}
\usepackage{enumitem}
\usepackage{graphicx}
\usepackage{cite}
\usepackage{color}
\usepackage{amsmath,amssymb,amsthm}
\usepackage{mathrsfs}
\usepackage{slashed}
\usepackage[english]{babel}
\usepackage[left=1cm,top=1cm,bottom=1cm,right=1cm,nohead,nofoot]{geometry}

\def \be {\begin{equation}}
\def \ee {\end{equation}}
\def \bea {\begin{eqnarray}}
\def \eea {\end{eqnarray}}
\def \nn {\nonumber}

\def \rr {\raise.35ex\hbox{\small $\prime$}\kern-.17em{\mbox{\large $\imath$}}}

\def \dels {\partial\kern-.6em /\kern.1em}
\def \As {{A\kern-.5em / \kern.5em}}
\def \Ds {D\kern-.7em / \kern.5em}

\def \ks {k\kern-.5em /}
\def \ls {l\kern-.5em /}

\newcommand{\ci}[1]{}

\newcommand{\ke}{\rangle}

\newcommand{\lb}{\left(}
\newcommand{\rb}{\right)}

\newcommand{\ba}{\begin{eqnarray}}
\newcommand{\ea}{\end{eqnarray}}
\newcommand{\bal}{\begin{align}}
\newcommand{\eal}{\end{align}}
\newcommand{\bay}[1]{\left(\begin{array}{#1}}
\newcommand{\eay}{\end{array}\right)}

\newcommand{\st}[1]{|#1\ke}

\def\xe{{\epsilon}}

\reversemarginpar
\newcommand{\hide}[1]{}

\makeatletter

\newlist{axioms}{enumerate}{2}
\setlist[axioms,1]{label=\textbf{A\arabic{axiomsi}.}, ref=A\arabic{axiomsi}}
\setlist[axioms,2]{label=\textbf{A\arabic{axiomsi}\rlap{\myEnumCounter{axiomsii}}.},%
                   ref=A\arabic{axiomsi}\myEnumCounter{axiomsii},%
                   align=parleft,%
                   leftmargin=0em,%
                   itemsep=1.4ex,%
                   before={\stepcounter{axiomsi}}}

\usepackage{tikz,pgf}
\usetikzlibrary{shapes}
\usetikzlibrary{calc}
\usetikzlibrary{decorations.pathmorphing}
\usetikzlibrary{decorations.pathreplacing,shapes.misc}
\usetikzlibrary{positioning}
\usetikzlibrary{arrows}
\usetikzlibrary{decorations.markings}

\tikzset{snake it/.style={decorate,decoration={snake,segment length=1.5mm, amplitude=.3mm}}}
\tikzset{biggerarrow/.style={
    decoration={markings,mark=at position 1 with {\arrow[scale=1.5]{>}}},
    postaction={decorate},
    shorten >=0.4pt}}
\tikzset{arrow at middle/.style={decoration={
    markings,
    mark=at position 0.5 with {\arrow{>}}}}}

\begin{document}

\begin{titlepage}

\begin{center}

\hfill
\vskip .2in

\textbf{\LARGE
Theoretical Properties of Entropy in a Strong Coupling Region
\vskip.5cm
}

\vskip .5in
{\large
Chen-Te Ma$^a$ \footnote{e-mail address: yefgst@gmail.com}\\
\vskip 3mm
}
{\sl
${}^a$
Department of Physics and Center for Theoretical Sciences, \\
National Taiwan University, Taipei 10617, Taiwan, R.O.C.
}\\
\vskip 3mm
\vspace{60pt}
\end{center}
\newpage
\begin{abstract}
Entropy is a quantity for counting physical degrees of freedom in a system. At a finite temperature, one can use thermal entropy to study thermodynamical properties. At zero temperature, entanglement entropy is expected to provide a suitable order parameter of a phase structure. Especially, the entanglement entropy exhibits an interesting codimension two area law in a strongly coupled conformal field theory. We compute thermal entropy in a non-relativistic model with an infinite fermion mass limit from an exact effective potential to obtain thermal entropy at an infinite strong coupling limit. The computational result provides vanishing thermal entropy at an infinite strong coupling limit with a finite lattice spacing. The non-trivial topological term can be included in the strongly coupled lattice system to obtain the non-trivial entropy and the topology can be marked from the entropy. We first compute the thermal entropy in a two dimensional lattice topological quantum field theory to study the lattice artifact and also argue that a theory possibly has translational invariance if a system does not have a volume law in the entanglement entropy. Finally, we show that a coefficient of a universal term of the entanglement entropy should not be affected by a choice of an entangling surface in two dimensional conformal field theory for one interval case. We also discuss a choice of the entangling surface in the entanglement entropy in two dimensional $\mathrm{CP}^{N-1}$ model at the large $N$ limit.
\end{abstract}

\end{titlepage}

\section{Introduction}
\label{1}
Quantum gravity theory is expected to combine the general relativity and the quantum mechanics and obey holographic principle. The holographic principle states that degrees of freedom in a system are encoded in the boundary of the system. The physical degrees of freedom of the thermodynamical system is proportional to the volume of the system. The holographic principle should restrict our construction of quantum gravity. A candidate of a perturbative quantum gravity theory is string theory. The string theory gives a conjecture of the holographic principle from anti-de Sitter/Conformal field theory (AdS/CFT) correspondence. The AdS$_d$/CFT$_{d-1}$ correspondence conjectures that equivalence of the Hilbert spaces between a $d$ dimensional weakly coupled bulk theory with the AdS background and a $d-1$ dimensional strongly coupled conformal field theory.

We first introduce the entanglement in quantum field theory and the problems of the gauge field in the entanglement. The important model in a strong coupling region is the quantum chromodynamics (QCD) model. The model describes dynamics of quarks and gluons and was also confirmed by experiments. The gauge sector of the QCD model is the Yang-Mills gauge theory with the non-Abelian grauge group SU(3). To study the quantum entanglement, which is a physical phenomenon for that a quantum state cannot be factorized, in the Yang-Mills gauge theory, we need to redefine the entanglement entropy. A universal term of the entanglement entropy in the Yang-Mills gauge theory suffers from a gauge invariant issue or decomposition problems, which provides the negative contribution to the entanglement entropy \cite{Donnelly:2012st}, from the spatial Wilson loops \cite{Casini:2013rba}. To define a gauge invariant entanglement entropy, one used a choice of centers, which commutes to other operators in the Hilbert space, or a choice of entangling surfaces to compute the entanglement entropy \cite{Casini:2013rba} in a gauge heory and also more generic quantum field theory {\cite{Ma:2015xes}. If one considers the QCD model in a strongly coupled region, the entanglement entropy should vanish because one should expect that the QCD model is the color singlet in the strong coupling region. Therefore, one should expect that the decomposition problems in the QCD model are not problematic in a strongly coupled region. The entanglement entropy was exactly computed in the two dimensional Yang-Mills gauge theory \cite{Gromov:2014kia}.

  The other point of view in the decomposition problems and a strongly coupled region is that the factorization problems of the entanglement entropy occur due to an ultraviolet (UV) scale \cite{Harlow:2015lma}. The large $N$ or a weakly coupled CP$^{N-1}$ theory on a two dimensional lattice does not suffer from the factorization problems and the model in the AdS background under the limit should be dual to a strongly coupled conformal field theory \cite{Kim:2016ipt}.  

The entanglement entropy of a subregion in the strongly coupled CFT is interesting in the theoretical physics because one can use the AdS geometry to obtain the exact solution of the entanglement entropy in the CFT. The entanglement entropy of CFT$_2$ for $N$ intervals was already provided from a geometric way \cite{Faulkner:2013yia}. Many exact understanding of the entanglement entropy were found in CFT$_2$. The entanglement entropy of one interval \cite{Casini:2004bw} and two disjoint intervals \cite{Calabrese:2010he} in CFT$_2$ were already computed. One can set different boundary conditions on an entangling surface to consider entanglement entropy with centers in CFT$_2$ \cite{Ohmori:2014eia} and used the same way to consider entanglement with centers to demonstrate that the mutual information is independent of a choice of centers \cite{Huang:2016bkp}, not same as in the entanglement entropy. Because the entanglement entropy is hard to study exactly in generic quantum field theory, the understanding of the entanglement entropy should rely on the numerical study. For example, Sachdev-Ye-Kitaev (SYK) model \cite{Fu:2016yrv}. If one can use the AdS geometry or the holographic way to easily obtain the exact solution of the entanglement entropy, it should be interesting. The holographic entanglement entropy can be clearly understood from the conformal mapping on the spherical entangling surface in the quantum field theory \cite{Casini:2010kt} or the AdS geometry \cite{Casini:2011kv}. The entanglement entropy with the trivial center also satisfies the subadditivity \cite{Araki:1970ba} and the strong subadditivity \cite{Lieb:1973cp} for all reduced density matrix. The mathematical proof is quite trouble, but the holographic way can easily show these two inequalities. Thus, the holographic entanglement entropy should have concrete evidences and be useful now \cite{Casini:2011kv}. Other related holographic studies are that the scale invariant field theory in four dimensions implies the conformal field theory in four dimensions in absence of a dimension two scalar operator \cite{Naseh:2016maw}, holographic entanglement thermodynamics \cite{Allahbakhshi:2013rda} and the entanglement entropy in the ${\cal N}=4$, where ${\cal N}$ is a number of the supercharges, supersymmetric Yang-Mills theory \cite{Huang:2014pda}. 

The above discussion is the entanglement in the continuum field theory. Now we discuss the entanglement in the lattice field theory. The lattice field theory is useful for obtaining numerical solution of the continuum field theory. However, the simulation is very hard usually. The interesting lattice QCD model suffers from the sign problem in the finite density case \cite{Braun:2012ww}. Many people are interested in using the mean-field theory \cite{Chen:2003vy} and studying the topological charge \cite{Bruckmann:2009cv} in the two dimensional lattice CP$^{N-1}$ model with a theta term \cite{Plefka:1996ks} to understand how to overcome the sign problem because the sign problem of the QCD model is hard to solve exactly now. 

The entanglement entropy in the lattice gauge theory is hard to study because one also needs to be careful about the decomposition problem. One proposed the extended lattice construction to study the entanglement entropy in the lattice gague theory \cite{Buividovich:2008kq}. One proved that the extended lattice construction, which enlarges the Hilbert space, is equivalent to the electric choice of centers \cite{Casini:2013rba}. The simulation in the four dimensional lattice Abelian gauge theory showed that the mutual information is not proportional to the center charge and the strong subadditivity can be violated on the lattice \cite{Casini:2014aia}, which was also confirmed from other lattice model \cite{VanAcoleyen:2015ccp}. Because the center charges in four dimensional non-interacting field theory can be related to the universal term of the entanglement entropy in the ${\cal N}$=4 U($N$) supersymmetric Ynag-Mills theory, the numerical study in the mutual information and the strong subadditivity should provide physical insight to a strongly coupled system.

Although the entanglement entropy is hard to compute, one can use the strong coupling expansion to obtain the entanglement entropy in the lattice U($N$) Yang-Mills gauge theory \cite{Ma:2015xes}. Note that the strong coupling expansion in the lattice gauge theory possibly does not provide a physical study at the continuum limit. The entanglement entropy in the lattice U($N$) Yang-Mills gauge theory still obeys the codimension two area law as in the holographic entanglement entropy, but it vanishes when the coupling constant goes to infinity \cite{Ma:2015xes}. The entanglement entropy in the ${\cal N}=4$ supersymmetric Yang-Mills theory does not vanish in the strong coupling region \cite{Huang:2014pda}. It is interesting to find that the entanglement entropies have the different behaviors in the strong coupling region. We will compute the entanglement entropy in a lattice theory to demonstrate the same behavior as in the entanglement entropy in the lattice U($N$) Yang-Mills gauge theory at the strong coupling limit.

The entanglement entropy is divergent in quantum field theory usually so it should depend on the regularization parameters. When one considers a theory with a finite dimensional Hilbert space, the entanglement entropy can be finite. One assumed that a theory has translational invariance, Poincaré symmetry, causality and finite entanglement entropy, then the entanglement entropy was determined from non-negative constant area terms \cite{Casini:2003ix}. The result is exact and still provided the codimension two area law as in the bulk minimum area of the holographic entanglement entropy.

Our goal of this paper is to study the entanglement entropy in a strongly coupled regime. We first consider a non-relativistic four fermion interaction with a spin imbalance at an infinite strong coupling limit and an infinite fermion mass limit to obtain vanishing thermal entropy. This result provides a supporting evidence for that the entanglement entropy vanishes in a strongly coupled lattice system if interacting terms between different sites are absent under some limits. The result is consistent with the strong coupling expansion in the lattice SU($N$) Yamg-Mills gauge theory. Here we also show that the thermal entropy also vanishes at the strong coupling limit. This shows no physical degrees of freedom in the strongly coupled regime in this non-relativistic fermion model. In this non-relativistic fermion model, the momentum cut-off is finite. When the momentum cut-off is finite and the coupling constant goes to infinity, the kinetic term should be truncated and only local interacting term survives in this non-relativistic fermion model. If one first considers the infinite momentum cut-off or the continuum field theory, the kinetic term may not be truncated. Thus, we demonstrate this thing clearly in this non-relativistic fermion model. This result is also consistent with the strong coupling expansion in the lattice gauge theory.

To study the entanglement entropy in a strong coupling region, the lattice construction is necessary. Even if we lose a kinetic term of a lattice field theory under some limits, we can include topological terms to obtain the non-trivial entropy. The topology can also be demonstrated by the entropy in the strongly coupled lattice system when the thermal entropy vanishes in the strongly coupled lattice system.

We also discuss the entanglement entropy in the two dimensional Einstein-Hilbert gravity theory. The entanglement entropy in a gravity theory does not have a clear definition. Thus, we directly use a replica trick or an $n$-sheet manifold to give an operational definition for the entanglement entropy because we only consider the compact manifolds in the two dimensional Einstein-Hilbert gravity theory.  The replica trick in the two dimensional Einstein gravity theory sums over all Riemann surfaces and the result also obtains $\alpha+m\beta$, where $m$ is a number of intervals and $\alpha$ and $\beta$ are constants, as in the two dimensional finite entropy \cite{Casini:2003ix}. Then we also argue that the translational invariance rules out a volume law of the entanglement entropy at zero temperature in an infinite size system without mass scales, except for a cut-off. We expect that a limit of vanishing cut-off should imply that physical quantities should be independent of the cut-off. Finally, we discuss whether universal terms of the entanglement entropy in two dimensional conformal field theory depend on a choice of entangling surfaces. In two dimensional conformal field theory, the known geometric method \cite{Faulkner:2013yia} provided that the universal contribution of the entanglement entropy comes from the bulk geometry. The bulk geometry should not be modified from a choice of entangling surfaces. We also use a symmetry principle \cite{Casini:2004bw} and unique mutual information \cite{Huang:2016bkp} to show that the coefficient of the universal term of the entanglement entropy for single interval should be independent of a choice of entangling surfaces \cite{Ohmori:2014eia}. Hence, we provide the consistent study between the bulk theory and the boundary theory. In a two dimensional bulk theory or a CP$^{N-1}$ model at the large $N$ limit, this theory is also expected to have a holographic duality \cite{Kim:2016ipt} because it approaches to a non-interacting scalar field theory. Therefore, a choice of entangling surfaces in the CP$^{N-1}$ model possibly does not affect entanglement entropy. 

We first compute thermal entropy in the non-relativistic model with a four fermion interaction and a spin imbalance in Sec.~\ref{2} and discuss thermal entropy in two dimensional topological lattice models in Sec.~\ref{3}. Then we discuss entanglement entropy in the two dimensional Einstein-Hilbert gravity theory, in which a measure of the theory is defined by a metric field \cite{Ma:2016deg}, in Sec.~\ref{4}. We also discuss dependence of a choice of entangling surfaces in entanglement entropy and two dimensional conformal field theory \cite{Ma:2016deg} in Sec.~\ref{5}. Finally, we give our conclusion in Sec.~\ref{6}.

\section{Non-Relativistic Fermion Theory}
\label{2}
We consider a non-relativistic fermion theory with a four fermion interaction and spin imbalance on a four dimensional lattice to compute the thermal entropy from the effective potential exactly when we take an infinitely fermion mass limit. The result provides that the thermal entropy vanishes at an infinite strong coupling limit, consistent with the result of the strong coupling expansion of the lattice U($N$) Yang-Mills gauge theory \cite{Ma:2015xes}.
\subsection{Action}
We begin our discussion from the continuum Euclidean action
\bea
S_{\mathrm{FF}}=\int d^4x\ \bigg\lbrack\psi^{\dagger}\bigg(\partial_{\tau}-\frac{\nabla^2}{2M}-\mu\bigg)\psi-\frac{1}{2m^2}(\psi^{\dagger}\psi)^2\bigg\rbrack,
\eea
where $M$ is the fermion mass, $m^2$ controls strength of interaction:
\bea
\psi\equiv
\left(\begin{array}{c}
\psi_{\uparrow} \\
\psi_{\downarrow}
\end{array}\right)\,, \qquad
\mu\equiv
\left(\begin{array}{cc}
\mu_{\uparrow}& 0\\
0&\mu_{\downarrow}
\end{array}\right)\,,
\eea
and the spin imbalance is controlled by different values of the up and down chemical potentials \big($\mu_{\uparrow}$ and $\mu_{\downarrow}$\big). To compute the effective potential exactly, we introduce an auxiliary field $\phi$ as in the following action
\bea
S_{\mathrm{FF1}}=\int d^4x\ \bigg\lbrack\psi^{\dagger}\bigg(\partial_{\tau}-\frac{\nabla^2}{2M}-\mu\bigg)\psi+\frac{m^2}{2}\phi^2-\psi^{\dagger}\phi\psi\bigg\rbrack.
\eea
The thermal entropy can be computed when we take an infinite fermion mass limit. The lattice action under the limit is given by 
\bea
S_{\mathrm{FFL}}=\sum_n\psi^{\dagger}_n\bigg(\psi_n-\exp(\mu)(1+\phi_n)\psi_{n-\hat{e}_0}\bigg)+\frac{m^2}{2}\phi_n^2.
\eea
Because we work in the case of a finite lattice spacing or a finite momentum cut-off, the computation under the infinite fermion mass limit is well-defined.
The partition function of the lattice theory \cite{Braun:2012ww, Chen:2003vy} is given by
\bea
Z_{\mathrm{FFL}}=\int D\phi\ \exp\bigg(-\frac{m^2}{2}\sum_n\phi_n^2\bigg)\det\bigg(\tilde{K}(\mu_{\uparrow})^T\tilde{K}(\mu_{\downarrow})\bigg),
\eea
where
\bea
(K\psi)_n=\psi_n-\exp(\mu)(1+\phi_n)\psi_{n-\hat{e}_0},
\eea
$\tilde{K}$ only acts on the coordinates space (or it does not act on spinor indices), $\hat{e}_0$ is a unit vector of the time direction, and the integers are labeled by the indices $i$-$z$.
\subsection{Thermal Entropy}
The exact solutions on a lattice are hard to obtain usually, but if we take an infinitely heavy fermion mass limit, the spatial derivative terms vanish to simplify our computation. This result is also the same as in the result from the strong coupling expansion of the lattice U($N$) Yang-Mills gauge theory. If we have two regions $A$ and $B$ on a lattice and want to obtain the entropy in the region $A$, we need to compute the partition function of the $n$-sheet manifold \cite{Buividovich:2008kq}. The period of the region $A$ is $nN_{\tau}$ and the region $B$ is $N_{\tau}$, where $N_{\tau}$ is a number of lattice size of the time direction. We also have cuts at the time slices $t=kN_{\tau}$, where $k=0, 1, \cdots, n$, in the region $B$. Regions $A$ and $B$ share the same boundary. In this set-up, we find that the partition function of the region $A$ and the partition function of the region $B$ can be computed separately because we do not have spatial derivative terms. The partition function of the region $B$ does not contribute to the entanglement entropy of the region $A$. Hence, we only need to compute the partition function in the region $A$. Because we only consider mean field level, we ignore quantum fluctuation of $\phi_n$. We also have
\bea
\ln\det A =\mbox{Tr}\ln A=\sum_p \ln A_p,
\eea 
where $A$ is a hermitian matrix and $p$ is momenta. Therefore, we can compute determinant of a matrix from the effective potential. The effective potential is given by
\bea
&&V_{\mathrm{eff}}
\nn\\
&=&\frac{m^2}{2}\phi^2-\frac{1}{N_s^3N_{\tau}}\sum_p\ln\bigg(1-\exp(\mu_{\uparrow})(1+\phi)\exp(iE)\bigg)
\nn\\
&&-\frac{1}{N_s^3N_{\tau}}\sum_p\ln\bigg(1-\exp(\mu_{\downarrow})(1+\phi)\exp(-iE)\bigg),
\eea
where $N_s$ is a number of the lattice size of the spatial directions and $E$ or $p_0$ is the energy. Now we compute the second term of the effective potential as the following computations:
 \bea
&&\frac{\delta V_{\mathrm{eff}}}{\delta\exp(\mu_{\uparrow})}
\nn\\
&=&\frac{1}{N_s^3N_{\tau}}\sum_p\frac{\exp(iE)}{1-\exp(\mu_{\uparrow})(1+\phi)\exp(iE)}\nn\\
&=&\frac{1}{N_s^3N_{\tau}}\sum_p\frac{1}{\exp(-iE)-\exp(\mu_{\uparrow})(1+\phi)}\nn\\
&=&\frac{1}{N_s^3N_{\tau}}\sum_{\vec{p}}\oint_C dE\ 
\nn\\
&&\times\frac{1}{\big(\exp(-iE)-\exp(\mu_{\uparrow})(1+\phi)\big)\big(\exp(iEN_{\tau})+1\big)}\frac{N_{\tau}}{2\pi}\nn\\
&=&\sum_{\vec{p}}\frac{1}{N_s^3N_{\tau}}\frac{N_{\tau}}{2\pi}(-2\pi i)
\nn\\
&&\times\frac{1}{-i\exp(\mu_{\uparrow})(1+\phi)\bigg(\big(\exp(\mu_{\uparrow})(1+\phi)\big)^{-N_{\tau}}+1\bigg)}\nn\\
&=&-\frac{\big(\exp(\mu_{\uparrow})(1+\phi)\big)^{N_{\tau}-1}}{1+\big(\exp(\mu_{\uparrow})(1+\phi)\big)^{N_{\tau}}}.\nn
\eea
Because we consider the anti-periodic boundary condition of the fermion field, we have the below conditions: 
\bea
&&\exp(ip_iN^i)=-1, \qquad p_i=\frac{2\pi(n_i+\frac{1}{2})}{N_i}, 
\nn\\
&&n_i=-\bigg\lbrack\frac{N_i}{2}\bigg\rbrack, \cdots, 0, \cdots, \bigg\lbrack\frac{N_i-1}{2}\bigg\rbrack,
\eea
where $[X]$ is the maximum interger and is liter than $X$ or equals $X$.
Hence, we sum over all energy modes, which is equivalent to doing the complex integration with the contour $C$, which encloses the poles of the function $\exp(iEN_{\tau})+1=0$. If our contour encloses all poles of the integrand, the integration should vanish. Thus, we obtain the fourth equality from the other pole. We do the integration to obtain the effective potential. The third term of the effective potential can be computed similarly. Thus, the effective potential is given by
\bea
&&V_{\mathrm{eff}}
\nn\\
&=&\frac{m^2}{2}\phi^2
-\frac{1}{N_{\tau}}
\ln\bigg\lbrack\bigg(1+\big(\exp(\mu_{\uparrow})(1+\phi)\big)^{N_{\tau}}\bigg)
\bigg(1+\big(\exp(\mu_{\downarrow})(1+\phi)\big)^{N_{\tau}}\bigg)\bigg\rbrack.
\nn\\
\eea
If we consider the $n$-sheet manifold, the effective potential becomes
\bea
&&V_{\mathrm{eff}}
\nn\\
&=&
\frac{m^2}{2}\phi^2
\nn\\
&&-\frac{1}{nN_{\tau}}
\ln\bigg\lbrack\bigg(1+\big(\exp(\mu_{\uparrow})(1+\phi)\big)^{nN_{\tau}}\bigg)
\bigg(1+\big(\exp(\mu_{\downarrow})(1+\phi)\big)^{nN_{\tau}}\bigg)\bigg\rbrack.
\nn\\
\eea
The thermal entropy is given by
\bea
S_{\mathrm{TE}}&=&\lim_{n\rightarrow 1}\bigg(-\frac{\partial}{\partial n}\ln\frac{Z_n}{Z^n}\bigg)
\nn\\
&=&N_s^3\ln\big((1+\xi_1^{N_{\tau}})(1+\xi_2^{N_{\tau}})\big)
-N_s^3N_{\tau}\frac{\ln\xi_1}{1+\xi_1^{-N_{\tau}}}-N_s^3N_{\tau}\frac{\ln\xi_2}{1+\xi_2^{-N_{\tau}}},
\eea
where $Z_n$ is the partition function of the $n$-sheet manifold and $Z$ is the original partition function, $\xi_1\equiv\exp(\mu_{\uparrow})(1+\phi)$ and $\xi_2\equiv\exp(\mu_{\downarrow})(1+\phi)$, and $\phi$ satisfies the following equation
\bea
m^2\phi-\frac{1}{1+\xi_1^{-N_{\tau}}}\frac{1}{1+\phi}-\frac{1}{1+\xi_2^{-N_{\tau}}}\frac{1}{1+\phi}=0.
\eea
When we take the strong coupling limit ($m\rightarrow 0$), $\phi\rightarrow\infty$ and $m^2\phi\rightarrow 0$, the thermal entropy vanishes under the limit at each temperature. Thus, non-trivial entropy in this model only comes from thermal entropy, which gives a volume law. Because we do not have the spatial derivative terms under the limit, the result for the vanishing entanglement entropy is not non-trivial. The non-trivial point of this computation is that the thermal entropy also vanishes at the infinite strong coupling limit. The physical reason possibly be that the derivative term can be ignored on a finite lattice at the strong coupling limit because we only have finite momentum cut-off. Hence, the on-site interacting term becomes dominant. If we take the continuum limit first, we expect that the result possibly be different.

  This result also indicates that we do not have any physical degrees of freedom in the system under the limit. Hence, this suggests that non-trivial physical degrees of freedom in a lattice theory at the strong coupling limit is necessary to include topological terms. When the strongly coupled system with the trivial topology has the vanishing thermal entropy, the topology can be determined by the entropy in the strongly coupled regime.
  
Before we  move to the next section, we discuss more about the entanglement entropy in the strong coupling region. The most interesting system in the strong coupling region is the ${\cal N}=4$ supersymmetric Yang-Mills theory in four dimensions. The entanglement entropy does not vanish in the strong coupling region. This result seems to contradict the above conclusion. We remind that the above discussion is restricted to a lattice system, not a continuum field theory. The ${\cal N}=4$ supersymmetric Yang-Mills theory is a continuum field theory. Thus, our discussion wants to address on that the entanglement entropy should not have the same behavior between the continuum theory and  the lattice theory generically. One also find that the entanglement entropy in the lattice Yang-Mills gauge theory with the U($N$) group also vanishes in the strong coupling region \cite{Ma:2015xes}. Thus, the lattice Yang-Mills gauge theory and the non-relativistic lattice fermion model can show different behavior of the entanglement entropy from the ${\cal N}=4$ supersymmetric Yang-Mills in the strong coupling region. When we take the strong coupling limit, the non-relativistic lattice fermion theory only has the finite momentum cut-off, but the continuum theory can have the infinite momentum cut-off. We expect that some differences in the entanglement entropy possibly come from the momentum cut-off.

\section{Thermal Entropy in Two Dimensional Topological Field Theory}
\label{3}

Entropy of topological quantum field theory only depends on topology of a manifold. On a lattice, the entropy possibly vanishes at an infinitely strong coupling constant limit on a lattice, except for topological terms. Hence, the entropy in the strongly coupled lattice system can demonstrate topology. We consider the action of the topological theory
\bea
S_{\theta}=-\frac{i\theta}{2\pi}\int d^2x F_{01}.
\eea 

The lattice action of $S_{\theta}$ is given by
\bea
S_{p}=-\frac{\theta}{2\pi}\sum_p\ln\big(U_p\big),
\eea
where $U_p$ is the product of the link variables $U_{\mu}\equiv\exp(iA_{\mu})$ around a plaquette, by using a plaquette method. The lattice model needs to use the periodic boundary condition to preserve the topological property on a lattice. We compute the thermal entropy on a lattice from the plaquette method. We first define $U_p\equiv\exp(if_{\mu\nu})$, where $-\pi< f_{\mu\nu}\le\pi$. The partition function is given by \cite{Plefka:1996ks}:
\bea
Z_{p}&=&\prod_{\mu\nu}\int_{-\pi}^{\pi}\frac{df_{\mu\nu}}{2\pi}\exp\bigg(\frac{i\theta}{2\pi}\sum_{\rho\sigma}f_{\rho\sigma}\bigg)\sum_n\delta\bigg(\sum_{\delta\gamma}f_{\delta\gamma}-2\pi n\bigg)
\nn\\
&=&\prod_{\mu\nu}\int_{-\pi}^{\pi}\frac{df_{\mu\nu}}{2\pi}\exp\bigg(\frac{i\theta}{2\pi}\sum_{\rho\sigma}f_{\rho\sigma}\bigg)\sum_m\exp\bigg(im\sum_{\delta\gamma}f_{\delta\gamma}\bigg)
\nn\\
&=&\prod_{\mu\nu}\int_{-\pi}^{\pi}\frac{df_{\mu\nu}}{2\pi}\sum_m\exp\bigg(i\frac{\theta+2\pi m}{2\pi}\sum_{\rho\sigma}f_{\rho\sigma}\bigg)
\nn\\
&=&\sum_m\bigg\lbrack\frac{2}{\theta+2\pi m}\sin\bigg(\frac{\theta+2\pi m}{2}\bigg)\bigg\rbrack^V.
\eea
The thermal entropy is given by:
\bea
S_{\mathrm{TE}}&=&-\lim_{n\rightarrow 1}\frac{\partial}{\partial n}\frac{Z_n}{Z_1^n}
\nn\\
&=&-\lim_{n\rightarrow 1}\frac{\partial}{\partial n}\frac{\sum_m\bigg\lbrack\frac{2}{\theta+2\pi m}\sin\bigg(\frac{\theta+2\pi m}{2}\bigg)\bigg\rbrack^{nV}}{\bigg\lbrack\sum_i\bigg\lbrack\frac{2}{\theta+2\pi i}\sin\bigg(\frac{\theta+2\pi i}{2}\bigg)\bigg\rbrack^V\Bigg\rbrack^n}
\nn\\
&=&\ln\Bigg\lbrack\sum_m\bigg\lbrack\frac{2}{\theta+2\pi m}\sin\bigg(\frac{\theta+2\pi m}{2}\bigg)\bigg\rbrack^{V}\Bigg\rbrack
\nn\\
&&-\frac{V}{\sum_m\bigg\lbrack\frac{2}{\theta+2\pi m}\sin\bigg(\frac{\theta+2\pi m}{2}\bigg)\bigg\rbrack^{V}}
\nn\\
&&\times\sum_i\bigg\lbrack\frac{2}{\theta+2\pi i}\sin\bigg(\frac{\theta+2\pi i}{2}\bigg)\bigg\rbrack^{V}
\ln\bigg\lbrack\frac{2}{\theta+2\pi i}\sin\bigg(\frac{\theta+2\pi i}{2}\bigg)\bigg\rbrack
\eea
 and the thermal entropy can be rewritten as the classical Shannon entropy:
 \bea
S_{\mathrm{TE}}=-\lim_{n\rightarrow 1}\frac{\partial}{\partial n}\frac{Z_n}{Z_1^n}
=-\sum_m p_m\ln p_m,
 \eea
  where
\bea
p_m\equiv \frac{\bigg\lbrack\frac{2}{\theta+2\pi m}\sin\bigg(\frac{\theta+2\pi m}{2}\bigg)\bigg\rbrack^{V}}{\sum_i\bigg\lbrack\frac{2}{\theta+2\pi i}\sin\bigg(\frac{\theta+2\pi i}{2}\bigg)\bigg\rbrack^{V}}.
\eea

We find that thermal entropy in a two dimensional lattice topological quantum field theory can be rewritten as the classical Shannon entropy as in the thermal entropy of the one-form Abelian gauge theory \cite{Donnelly:2012st}. Thus,we show that this property is not modified by the lattice artifact in this lattice topological quantum field theory.

\section{Entanglement Entropy in Quantum Gravity Theory}
\label{4}
We discuss entanglement entropy in quantum gravity theory. Although quantum gravity theory has some expected properties, it is still hard to understand from the first principle. Thus, we first use the two dimensional Einstein-Hilbert action to compute the entanglement entropy. This theory is topological quantum theory and a conformal field theory, and it also appears in string theory to give different topology of a moduli space. Hence, the two dimensional Einstein-Hilbert theory is a suitable model to know theoretical properties of quantum gravity theory in different coupling regions. We assume that area law for the entanglement entropy, then we argue that the translational invariance is needed in a physical system.

\subsection{Two Dimensional Einstein-Hilbert Theory}
The action of the two dimensional Einstein-Hilbert theory is:
\bea
S_{EH}=-\frac{1}{16\pi G}\int d^2x\ \sqrt{\det{g_{\mu\nu}}}R=-\frac{1}{4G}\chi,
\eea
where $\chi\equiv 2-2g$, $g$ is a number of genus, $G$ is the two dimensional gravitational constant, and have the following terms:
\bea
R_{\mu\nu}&\equiv&\partial_{\delta}\Gamma^{\delta}_{\nu\mu}-\partial_{\nu}\Gamma^{\delta}_{\delta\mu}
+\Gamma^{\delta}_{\delta\lambda}\Gamma^{\lambda}_{\nu\mu}
-\Gamma^{\delta}_{\nu\lambda}\Gamma^{\lambda}_{\delta\mu}, 
\nn\\
 \Gamma^{\mu}_{\nu\delta}&\equiv&\frac{1}{2}g^{\mu\lambda}\bigg(\partial_{\delta}g_{\lambda\nu}+\partial_{\nu}g_{\lambda\delta}
-\partial_{\lambda}g_{\nu\delta}\bigg),
\nn\\
 R&\equiv& g^{\mu\nu}R_{\mu\nu}.
\eea 
We compute the entanglement entropy by an $n$-sheet method \cite{Gromov:2014kia} as the followings:
\bea
S_{\mathrm{EE}}&=&-\lim_{n\rightarrow 1}\frac{\partial}{\partial n}\frac{Z_n}{Z_1^n}
\nn\\
&=&-\lim_{n\rightarrow 1}\frac{\partial}{\partial n}\frac{\sum_{\chi^{\prime}}e^{\frac{1}{4G}\big(n\chi^{\prime}-2N(n-1)\big)}}{(\sum_{\chi}e^{\frac{\chi}{4G}}\big)^n}
\nn\\
&=&-\frac{\sum_{\chi^{\prime}}e^{\frac{1}{4G}\chi^{\prime}}\frac{1}{4G}(\chi^{\prime}-2N)}{\sum_{\chi}e^{\frac{\chi}{4G}}}
+\ln\bigg(\sum_{\chi}e^{\frac{\chi}{4G}}\bigg)+\cdots
\nn\\
&=&\ln\bigg(\sum_{\chi}e^{\frac{\chi}{4G}}\bigg)-\frac{1}{4G}\frac{\sum_{\chi^{\prime}}e^{\frac{\chi^{\prime}}{4G}}\chi^{\prime}}{\sum_{\chi}e^{\frac{\chi}{4G}}}
+\frac{\langle N\rangle}{2G}+\cdots,
\eea
where $2N$ is a number of ramification points, $\langle N\rangle$ is the expectation value of $N$ and $\chi$ is the Euler number.  The partition function is computed by summing over all Riemann surfaces (different numbers of the genus and different numbers of the ramification points). If we define the probability
\bea
p_i\equiv\frac{e^{\frac{1}{4G}\chi_i}}{\sum_{\chi}e^{\frac{1}{4G}\chi}}
\eea
in the classical Shannon entropy, we obtain the below result
\bea
-\sum_i p_i\ln p_i=\ln\bigg(\sum_{\chi}e^{\frac{\chi}{4G}}\bigg)-\frac{1}{4G}\frac{\sum_{\chi^{\prime}}e^{\frac{\chi^{\prime}}{4G}}\chi^{\prime}}{\sum_{\chi}e^{\frac{\chi}{4G}}}.
\eea
Thus. the entanglement entropy can be rewritten as
\bea
S_{\mathrm{EE}}=-\sum_ip_i\ln p_i+\frac{\langle N\rangle}{2G}+\cdots\ .
\eea
Indeed, the expression is very interesting because the result contains the classical entanglement and quantum entanglement and it is also quantum extension of the finite entropy \cite{Casini:2003ix}, in which they find that a two dimensional finite entropy has the same and unique form if a theory has translational invariance, Poincaré symmetry and causality. If we only consider a classical background and a spherical geometry, a number of ramification points correspond to a number of intervals because the two dimensional Einstein-Hilbert action also has the conformal symmetry \cite{Casini:2011kv}. We also get the consistent result $\alpha+\beta N$, where $\alpha$ and $\beta$ are constants, as in the two dimensional finite entropy \cite{Casini:2003ix}. For two dimensional field theories, the entangling surface is a point. Thus, we think that an analogous area quantity is defined by a number of ramification points and each interval has two ramification points. The quantum entanglement comes from the term
\bea
\frac{2\langle N\rangle}{4G}.
\eea
 The $\cdots$ in the entanglement entropy of the two dimensional Einstein-Hilbert theory comes from the dependence of $n$ or the degeneracy of closed manifolds.

In higher dimensions, the analogue term is given by
\bea
\frac{\langle A\rangle}{4G},
\eea
 where $A$ is a codimension two surface. This possibly motivates us to think that quantum gravity theory has the area operator 2$\hat{N}$ or $\hat{A}$. The sum over all ramification points in the path integral is also equivalent to summing over all classical configurations on an entangling surface. Hence, the result of the two dimensional Einstein-Hilbert action possibly inspires us to find that the entanglement entropy is the sum of a non-negative constant and a non-negative area term from physical principles or define a suitable area operator to explore quantum gravity theory. 
 
 The entanglement entropy in a gravity theory is hard to define or understand in a gravity theory. Here we used the replica trick to define the entanglement entropy in the two dimensional Einstein-Hilbert gravity theory because the two dimensional Einstein-Hilbert gravity theory is defined in the compact manifolds. The evidence is that we can reproduce the result of the two dimensional finite entropy. Now we give the other evidence to the definition of the entanglement entropy, the $n$-sheet manifold. We know that the two dimensional dilaton gravity theory have the holographic boundary theory, Sachdev-Ye-Kitaev (SYK) model \cite{Fu:2016yrv}. When the dilaton field is a constant, the two dimensional dilaton gravity theory should become the two dimensional Einstein-Hilbert gravity theory. The entanglement entropy of the SYK model is proportional to the sites of a subsystem when the SYK model has enough Majorana fermion fields \cite{Fu:2016yrv}. The number of ramification points, $N$, corresponds to the number of intervals in the two dimensional Einstein-Hilbert gravity theory. The sites of a subsystem and the number of ramification can be seen as physical degrees of freedom in a subsystem. Thus, we argue that the definition of the entanglement entropy is suitable in the two dimensional Einstein-Hilbert gravity theory.

\subsection{Non-Volume Law of the Entanglement Entropy}
We want to use dimensional analysis to argue that the entanglement entropy is impossible to have a volume law when a theory is translational invariant, and satisfies subadditivity at zero temperature in an infinite size system without mass scales, except for a cut-off. We expect that a physical quantity is independent of the cut-off when the cut-off approaches to zero. We first use the translational invariance, then the entanglement entropy must depend on translational invariant quantities. For example, size of a system. We also have subadditivity law \cite{Ma:2015xes} as the inequality
\bea
S_A+S_B\ge S_{AB}
\eea
if $\rho_{AB}$ is a density matrix in the Hilbert space of the regions $A$ and $B$, $H_{AB}$, isomorphic to the Hilbert space $\oplus_i H^i_A\otimes H^i_B$. Then we can show that the density of the entanglement entropy should be finite when a spatial volume goes to infinity as in the following proof:
\bea
\frac{S_A(k_1, k_2, \cdots, k_n)}{k_1k_2\cdots k_n}< \infty,
\eea
\bea
S_A(a_1, a_2, \cdots, a_n)
&=&S_A(p_1k_1+q_1, p_2k_2+q_2, \cdots, p_nk_n+q_n)
\nn\\
&\le& p_1p_2\cdots p_nS_A(k_1, k_2, \cdots, k_n)+(\mbox{low numbers of $p_i$}),
\nn\\
\eea
\bea
\frac{S_A(a_1, a_2, \cdots, a_n)}{a_1a_2\cdots a_n}
\le \frac{p_1 k_1}{a_1}\frac{p_2 k_2}{a_2}\cdots \frac{p_nk_n}{a_n}\frac{S_A(k_1, k_2, \cdots, k_n)}{k_1k_2\cdots k_n}+\cdots,
\eea
\bea
\lim_{a_1, a_2, \cdots a_n\rightarrow\infty}\frac{S_A(a_1, a_2, \cdots, a_n)}{a_1a_2\cdots a_n}\le
\frac{S_A(k_1, k_2, \cdots, k_n)}{k_1k_2\cdots k_n}< \infty,
\eea
in which we used $a_i\equiv p_ik_i+q_i$, $0\le q_i\le k_i-1$. Therefore, we obtain that the density of the entanglement entropy should be finite if a theory is translational invariant and satisfies the subadditivity law at zero temperature in a system with an infinite size without mass scales, except for a cut-off. We remind that the density of the entanglement entropy is a physical quantity only when the volume of the system goes to infinity. If our system is at zero temperature in an infinite size system without mass scales, except for a cut-off, we only have a regularization parameter or a cut-off and the side length of a sub-physical system with a unit of length. If we have a volume term in the entanglement entropy, the volume term should be proportional to $V/\epsilon^{D-1}$, where $V$ is the spatial volume of a subsystem, $\epsilon$ is the regularization parameter and $D$ is a number of the spacetime dimensions. Thus, it is easy to find that the density of the entanglement entropy should be divergent when we take the limit $\epsilon\rightarrow 0$. We argue that the non-volume law of the entanglement entropy at zero temperature in a system with an infinite size without mass scales, except for a cut-off, should need translational invariance and satisfy the subadditivity law, which can be shown from the time translational invariance in a quantum system. Because we expect that a perturbative quantum gravity theory also has the area law of the entanglement entropy as in the holographic entanglement entropy, our discussion possibly implies that the translational invariance is a necessary condition in a perturbative quantum gravity theory by ruing out the volume law of the entanglement entropy. Although we cannot include mass scales, except for a cut-off, in our discussion, a weakly coupled bulk gravity theory is expected to be described by the strongly coupled conformal field theory and the mass term breaks conformal symmetry or breaks scale invariance in a theory. Thus, our discussion possibly be generic when we discuss a perturbative quantum gravity theory.

\section{Two Dimensional Conformal Field Theories}
\label{5}
The codimension two minimum surface of a weakly coupled bulk gravity theory gives universal terms of the entanglement entropy in a strongly coupled conformal field theory. This means that the universal terms of the entanglement entropy in a strongly coupled conformal field theory can be rewritten in terms of geometrical quantities. A choice of an entangling surface in the entanglement entropy comes from a choice of decompositions of a Hilbert space, but the choice should not modify geometry of a bulk theory. Thus, we want to argue that a universal contribution of the entanglement entropy in a strongly coupled conformal field theory possibly does not depend on a choice of entangling surfaces or is unaffected by a choice of entangling surfaces. We first consider two dimensional conformal field theory. In this theory, we use mathematical methods to determine the coefficient of the universal term of the entanglement entropy for a single interval uniquely and this method can also be extended to some cases of multiple intervals. We discuss generic multiple intervals in two dimensional conformal field theory from a known geometric or holographic methods \cite{Faulkner:2013yia}. Finally, we consider an example, two dimensional CP$^{N-1}$ model at the large $N$ limit. The model approaches to a non-interacting scalar field theory under the large $N$ limit. Thus, we compute the entanglement entropy at the large $N$ limit and in the AdS background and have a unique form of the entanglement entropy for single interval.

\subsection{Two Dimensional Conformal Field Theory}
We first use translational invariance and strong subadditivity \cite{Ma:2015xes, Huang:2016bkp, Araki:1970ba, Casini:2014aia, Lieb:1973cp, VanAcoleyen:2015ccp}, then we obtain the inequality
\bea
S_A(l_A)+S_B(l_B)\ge S_{A\cup B}(l_{A\cup B})+S_{A\cap B}(l_{A\cap B}).
\eea
We consider a Poincaré symmetry or boost symmetry to constraint the length of the systems for one interval case in two dimensional quantum field theory. We have four points $a_1=(0, 0)$, $a_2=(0, c_1)$, $a_3=(d_2, c_1+d_1)$ and $a_4=(d_2+b_1, c_1+d_1)$, in which we use null coordinates $u=t+x$, $v=t-x$, to define the size $\sqrt{uv}$  of the systems $A$ ($\overrightarrow{a_1 a_3}=\sqrt{d_2(c_1+d_1)}\equiv l_A$), $B$ ($\overrightarrow{a_2a_4}=\sqrt{d_1(b_1+d_2)}\equiv l_B$), $A\cup B$ ($\overrightarrow{a_1a_4}=\sqrt{(d_2+b_1)(c_1+d_1)}\equiv l_{A\cup B}$) and $A\cap B$ ($\overrightarrow{a_2a_3}=\sqrt{d_1d_2}\equiv l_{A\cap B}$), and we assumed the coordinates $(u_i, v_i)\equiv a_i$. Thus, we obtain the below results: 
\bea
l_{A\cup B}\cdot l_{A\cap B}=l_{A}l_B=\sqrt{d_1d_2(d_2+b_1)(c_1+d_1)}, 
\qquad
 \frac{l_A}{l_{A\cup B}}=\frac{l_{A\cap B}}{l_B}\equiv\frac{1}{\lambda},
\eea
where $\lambda\ge 1$.
This interesting relation implies the inequalities:
\bea
S_A(l_A)-S_{A\cap B}(l_{A\cap B})\ge S_{A\cup B}(l_{A\cup B})-S_B(l_B)
=S_{A\cup B}(\lambda l_A)-S_B(\lambda l_{A\cap B}).
\nn\\
\eea
Finally, we use a modular transformation, SL(2, C) transformation, to restrict the form of the entanglement entropy:
\bea
x\rightarrow\frac{ax+b}{cx+d}, \qquad ad-bc=1,
\eea
which is a conformal transformation because scaling, inversion, and translation still preserves angles. The parameters $a$, $b$, $c$ and $d$ are complex constants. The below quantities are invariant under the modular transformation:
\bea
\frac{(u_2-u_3)(u_1-u_4)}{(u_1-u_3)(u_2-u_4)}, \qquad
 \frac{(v_2-v_3)(v_1-v_4)}{(v_1-v_3)(v_2-v_4)}.
\eea 
We also know that $F\equiv S_A(l_A)+S_B(l_B)-S_{A\cap B}(l_{A\cap B})-S_{A\cup B}(l_{A\cup B})$ is invariant under the modular transformation. From the equality
\bea
&&F\bigg(\frac{(u_2-u_3)(u_1-u_4)}{(u_1-u_3)(u_2-u_4)}, \frac{(v_2-v_3)(v_1-v_4)}{(v_1-v_3)(v_2-v_4)}\bigg)
\nn\\
&=&S_A\bigg(\sqrt{(u_1-u_3)(v_1-v_3)}\bigg)
+S_B\bigg(\sqrt{(u_2-u_4)(v_2-v_4)}\bigg)
\nn\\
&-&S_{A\cap B}\bigg(\sqrt{(u_2-u_3)(v_2-v_3)}\bigg)
-S_{A\cup B}\bigg(\sqrt{(u_1-u_4)(v_1-v_4)}\bigg),
\eea
we obtain the entanglement entropy of the region $A$
\bea
S_A=k_1\ln l_A+k_2,
\eea
where $k_1$ and $k_2$ are constants,
because we can derive $f=k_1\ln x+k_2$ for $g(x^2y^2)=f(x)+f(y)$ as the followings:
\bea
&&g(x^2y^2)=f(x)+f(y)\Rightarrow 2xy^2\frac{dg(x^2y^2)}{dx}=\frac{df(x)}{dx}
\nn\\
&&\Rightarrow \frac{2}{x}\frac{dg}{dx}(1)\equiv \frac{k_1}{x}=\frac{df(x)}{dx}
\nn\\
&&\Rightarrow f(x)=k_1\ln x+k_2.
\nn\\
\eea
Thus, the form of the entanglement entropy is determined from the translational invariance, the strong subadditivity, the boost symmetry and the conformal symmetry \cite{Casini:2004bw}. The determination of the entanglement entropy from mathematical methods can include different choices of entangling surfaces because we only impose the symmetry to constrain the form of the entanglement entropy and do not choose any choice of entangling surfaces. We cannot exclude that the coefficient of the universal term of the entanglement entropy is modified from a choice of entangling surfaces because we cannot restrict the value of the universal coefficient. To know that the dependence of a choice of entangling surfaces in the entanglement entropy, we need to know other entanglement quantities, which can be related to the coefficient of the universal term of the entanglement entropy. We already found an example to know that the mutual information does not depend on a choice of an entangling surface in two dimensional conformal field theory for multiple intervals \cite{Huang:2016bkp}. Now we show that the coefficient of the universal terms in the mutual information does not depend on a choice of entangling surfaces for one interval. Each region, $A$, $B$ and $ A\cap B$, is one interval. 

Now we show that the coefficient of the universal term in the entanglement entropy is not affect by a choice of entangling surfaces. If the entanglement entropy of the region $A$ is given by
\bea
S_A=k_1\ln l_A+k_2,
\eea
the mutual information is given by 
\bea
M=k_1\ln\frac{l_Al_B}{l_{A\cup B}}+k_2
\eea
if each region is single interval. We use the same method \cite{Huang:2016bkp} to show that the mutual information is independent of a choice of entangling surfaces and know that the parameter $k_1$ should not be affected by a choice of entangling surfaces. In other words, we remove operators on the boundary of the region $A\cup B$. Hence, the coefficient of the universal term for one interval in two dimensional conformal field theory should be unique. Now we mention the logic of showing uniqueness of the mutual information. We first insert boundary state in an entangling surface. Then the $n$-sheet partition function of a cylinder for considering ground state is given by
\begin{equation}
        Z_n=\bigg\langle a_1^{(n)}\bigg| \exp\left(\frac{\ell}{n} \frac{c}{12} \right) \bigg|a_2^{(n)}\bigg\rangle \label{eq:Zn}\,,
\end{equation}
where $\big|a_{1,2}^{(n)}\big\rangle$ are the boundary states from the cutoff circle $\ell =  \ln \lb
l_A/ \xe\rb^{2}$ and $c$ is the center charge. We can use the conformal mapping 
\be
w = \ln \frac {z - z_1}{z-z_2}\,,
\ee
to obtain the single interval from the cylinder. The R\'enyi entropy of the region $A$ for ground state in a single interval case is given by:
 \bea
        &&S_n
        \nn\\
         &\equiv&\frac{\ln\mathrm {Tr}\rho_A^n}{1-n}
        \nn\\
        &=&\bigg(1+\frac1n\bigg)\frac{c}{6}\bigg(\ln\frac{l_A}{\epsilon}\bigg) 
        \nn\\
        &&
+ \frac{1}{1-n}\bigg(s(a_1^{(n)})-ns(a_1^{(1)}) + s^*(a_2^{(n)})-ns^*(a_2^{(1)})\bigg),
        \nn\\
                \label{eq:subleadingRenyi}
\eea
where 
\bea
s\big(a_1^{(n)}\big) = \ln \langle {a_1^{(n)}}\st 0
\eea
 is the boundary entropy. Then we can find that the boundary entropy should disappear in the mutual information when we insert the boundary states to consider different choices of entangling surfaces. Therefore, we conclude that the coefficient of the universal term of the entanglement entropy must be $c/3$ for one interval without any modification from a choice of entangling surfaces. This method can also be extended to some cases of multiple intervals \cite{Huang:2016bkp}. From the known geometric construction \cite{Faulkner:2013yia}, the coefficient of the entanglement entropy for the multiple intervals can be determined by the AdS$_3$ geometry. Because we expect that geometry of a theory should not be modified by a choice of entangling surfaces, our study should be consistent with this geometric study. 
 
When we take regularization parameters be small and finite, the dominant term is a regularization dependent term and these terms are still finite. This study is consistent with the two dimensional finite entropy \cite{Casini:2003ix}. The consistency provides the evidences to the geometric study.

\subsection{Two Dimensional $\mathrm{CP}^{N-1}$ Model}
The action of the continuum $\mathrm{CP}^{N-1}$ model is
\bea
S_{\mathrm{cp}}=\beta N\int d^2x\
 \bigg(\partial_{\mu}z_i^*\partial_{\mu}z_i+\big(z_i^*\partial_{\mu}z_i\big)\big(z_j^*\partial_{\mu}z_j\big)\bigg),
\eea
where $z_i(x)$ is an $N$ component complex field satisfying $z_i^*z_i=1$, $\beta N\equiv 1/g$, and $g$ is a coupling constant. This model can approach to the non-interacting scalar field theory at the large $N$ limit. Thus, we can consider the bulk scalar field to find a holographic duality of one-dimensional strongly coupled conformal field theory \cite{Kim:2016ipt, Fu:2016yrv}. The non-interacting scalar field theory is also conformal field theory so we can compute the entanglement entropy when we take the limit $N\rightarrow\infty$ \cite{Casini:2010kt}. The entanglement entropy for one interval is also unique. The entanglement entropy in the AdS$_2$ case and large $N$ limit should be half of the entanglement entropy in the flat background, and proportional to $N$. 

Because the two dimensional $\mathrm{CP}^{N-1}$ model is also a conformal field theory when we consider the large $N$ limit $N\rightarrow\infty$, a coefficient of a universal term should be unique for any choice of entangling surfaces. Thus, we argue that the holographic entanglement entropy of  AdS$_2$/CFT$_1$ correspondence possibly does not have any dependence on a choice of entangling surfaces. 

Now we discuss a lattice $\mathrm{CP}^{N-1}$ model. The lattice model can be written by putting link variables as
\bea
S_{\mathrm{lcpg}}=-\beta N\sum_{x,\hat{\mu}}(z^*_{x+\hat{\mu}}\cdot z_x) U^*_{\mu}(x)
+(z^*_x\cdot z_{x+\hat{\mu}})U_{\mu}(x)
\eea
Thus, this seems that a choice of entangling surfaces may affect the entanglement entropy. We should remember that a choice of entangling surfaces comes from UV scale \cite{Harlow:2015lma}. The continuum limit in the lattice $\mathrm{CP}^{N-1}$ model is in the weak coupling region and under the large $N$ limit so the link variables becomes auxiliary fields. The non-dynamical fields are not problematic to give a non-gauge invariant universal term of the entanglement entropy \cite{Huang:2016bkp}. From the lattice point of view, we also get the consistent understanding with the continuum theory.

\section{Conclusion}
\label{6}
We studied various approaches or problems of the entropy in a strong coupling region. The first study is to consider the non-relativistic four fermion model with a spin imbalance on a lattice. We computed the thermal entropy by setting an infinite fermion mass and taking an infinite strong coupling constant to give vanishing thermal entropy. This means that we do not have any physical degrees of freedom in the non-relativistic fermion model in an infinite strong coupling region with an infinite fermion mass as in the results of a strong coupling expansion in the lattice U($N$) Yang-Mills gauge theory \cite{Ma:2015xes}. Thus, this implies that we need topological quantum field theory to obtain non-trivial physical degrees of freedom in strongly coupled lattice theories. It is also interesting to note that the entropy from the topological terms can indicate topology in the strongly coupled lattice system. Our interpretation is that the vanishing entropy in the strongly coupled lattice system possibly be due to the finite momentum cut-off in a lattice theory because the kinetic term in this non-relativistic fermion model should be truncated when the coupling constant of the local interacting term goes to infinite. We also showed the equivalence between the classical Shannon entropy and the thermal entropy in the two dimensional lattice theta term without suffering from the lattice artifact. 

 We also computed the entanglement entropy in the two dimensional Einstein gravity theory, which is also topological quantum field theory and conformal field theory, by summing over all different numbers of ramification points and genus. Then the entanglement entropy in the two dimensional Einstein gravity theory contains the classical Shannon entropy and the analogous codimension two surface term (proportional to the expectation values of a number of ramification points). Therefore, we expect that this study possibly implies that a perturbative quantum gravity theory needs the area operator and we strongly expect that the entanglement entropy in a perturbative quantum gravity theory should have the area law. We also use the translational invariance and subadditivity, which can be shown by the unitary or the time translational invariant, in an infinite size system at zero temperature without mass scales, except for a cut-off, to rule out the volume law in the entanglement entropy. Hence, this possibly indicates that a perturbative quantum gravity theory needs translational invariance. 
 
 Finally, we used symmetry principles \cite{Casini:2004bw} and uniqueness of the mutual information \cite{Huang:2016bkp} to show that a coefficient of the universal term of the entanglement entropy for the single interval case in two dimensional conformal field theory should be unique and the result can be extended to some cases of the multiple intervals. The known geometric method \cite{Faulkner:2013yia} can determine the coefficient of the universal terms of the entanglement entropy in two dimensional conformal field theory for the multiple intervals and we do not expect that the geometry of a bulk gravity theory can be modified by choosing different entangling surfaces. Thus, our study provides the consistent study between the three dimensional bulk geometry and two dimensional conformal field theory. We also discuss the dependence of a choice of entangling surfaces of the entanglement entropy in the two dimensional CP$^{N-1}$ theory and discuss its implication in the holographic duality. 
 
 The area law of the entanglement entropy gives a geometric way to understand the entanglement entropy. Now we studied various cases to understand the area law generically. We are especially interested in whether the area law vanishes in the strongly coupled quantum field theory. From our study, the area law possibly vanishes when we take a finite momentum cut-off or consider a finite lattice size and consider an infinite strong coupling constant. Hence, this direction may let us explore more about differences between the lattice theory and the continuum theory in the entanglement entropy at the strong coupling limit.
 
We usually expect that the mutual information is proportional to a center charge in conformal field theory. In the four dimensional Abelian gauge theory gives a counter example to us \cite{Casini:2014aia}. The mutual information is only proportional to a bulk central charge \cite{Casini:2014aia}. The boundary central charge is canceled in the mutual information \cite{Casini:2014aia}. The definition of the entanglement entropy in gauge theories needs to sacrifice quantum fluctuation of an entangling surface to give new information to boundaries or entangling surfaces \cite{Casini:2013rba}. We think that the strongly coupled conformal field theory only stores information of the bulk, then after flowing to a weakly coupled theory, some information of the bulk flow to the boundary. Thus, we do not see a boundary central charge in the mutual information. From the point of view of the holographic principle, we expect that a holographic dual theory \cite{Naseh:2016maw} needs to store informal in the bulk or on the boundary totally because we do not think that physics of all higher dimensions can be deduced from the lowest dimensional physics (one dimensional physics). Thus, this may give the interpretation why the strongly coupled conformal field theory should be dual to a weakly coupled bulk gravity theory. 

To prove uniqueness of the universal terms of the entanglement entropy in higher dimensional conformal field theory \cite{ Calabrese:2010he, Allahbakhshi:2013rda}, one cannot use the same method to show because the mutual information possibly does not count all center charges. The method may be extended to strongly coupled conformal field theory because one expects that the entanglement entropy in strongly coupled conformal field theory does not have a boundary center charge in the universal terms of the entanglement entropy. Thus, the mutual information may contain all central charges of the entanglement entropy in the strongly coupled conformal field theory. Our paper should give a starting point in this direction.
\section*{Acknowledgment}
The author would like to thank Horacio Casini, Daniel Harlow, Song He, Kazuo Hosomichi, Xing Huang, Yu-tin Huang and Ling-Yan Hung for their useful discussion. Especially, the author would like to thank Nan-Peng Ma for his suggestion and encouragement. The author thanks the Asia Pacific Center for Theoretical Physics at the Pohang University of Science and Technology, Yukawa Institute for Theoretical Physics at the Kyoto University, National Tsing Hua University, Tohoku University and Okinawa Institute of Science and Technology Graduate University. Discussions during the workshops ``Duality and Novel Geometry in M-theory'', ``Quantum Information in String Theory and Many-body Systems'', ``Novel Quantum States in Condensed Matter 2017'', ``The NCTS workshop on correlated quantum many-body systems: from topology to quantum criticality'', ``String-Math 2018'', ``Strings 2018'', ``New Frontiers in String Theory'' and ``Strings and Fields 2018'' were useful to complete this work.


\end{document}